# Challenges in the Theoretical Description of Nanoparticle Reactivity: Nano Zero-Valent Iron

**František Karlický, Michal Otyepka**

*Regional Centre of Advanced Technologies and Materials, Department of Physical Chemistry, Faculty of Science, Palacký University, 17. listopadu 1192/12, 771 46 Olomouc, Czech Republic*

Correspondence to: F. Karlický (frantisek.karlicky@upol.cz)

**ABSTRACT**

The reactivity of iron atoms, clusters and nanoparticles (nZVI) is of increasing interest owing to their important practical applications, ranging from the steel industry to water remediation technologies. Here, we provide an overview of computational methods and models that can be applied to study nZVI reactions and discuss their benefits and limitations. We also report current progress in calculations through recent examples treating the reactivity of nZVI particles. Finally, we consider the potential use of highly accurate methods with favorable scaling (such as quantum Monte Carlo or random phase approximation), which are currently considered too computationally expensive but are expected to become more amenable in the future as computer power increases.

## Introduction

In recent years, reductive technologies based on zero-valent iron (ZVI) and nanoscale zero-valent iron (nZVI, nanoFe$^0$) have become widely used for the decontamination of ground- and waste-water. The high reduction capacity of ZVI has been known for some time, but progress in the cheap and effective synthesis of micro- and particularly nano-scale ZVI has enabled reductive technologies to become effective and broadly applicable. Iron nanoparticles have been shown to perform well in the decontamination of various organic and inorganic pollutants.[1] Generally, nanoparticles of transition metals (TM) and noble metals are a significant component of modern nanotechnologies, where they can be used as highly effective catalysts or contribute to promising applications in energy, optics, electronics, drug delivery and medical diagnostics.

A detailed understanding of the mechanisms of nZVI reactions should make it possible to optimize reduction and catalytic processes and identify other potential applications of this material. In this respect, theoretical methods provide unique information with atomic resolution. However, the theoretical study of nZVI is complicated by the fact that it is difficult to balance model complexity and accuracy of the theoretical method. Because large systems such as the size of nZVI particles are computationally intractable, they are usually modeled as a cluster of iron atoms or as an infinite (periodic) surface. However, as computer power has gradually increased over the past few decades, it has become possible to model larger and larger clusters.

## Bridge between Experiments and Computations

Energy changes along the reaction pathway make a bridge between experiments and computations, as the energy differences between minima correspond to thermodynamic properties and barriers heights to kinetic properties. As here we focus on reactivity, the kinetics can be represented by the Eyring equation,

$$k = \kappa \left(\frac{k_B T}{h}\right) e^{-\Delta G^{\ddag}/RT}$$

which relates the (measured) rate constant $k$ to the (calculated) Gibbs free energy barrier $\Delta G^{\ddag}$. However, one should keep in mind potential limitations connected with application of Eyring equation for reactions of nanoparticles. Nevertheless, small changes in $\Delta G^{\ddag}$ induce large variations of the rate constant because of the exponential dependence of the rate constant on $\Delta G^{\ddag}$. This implies that activation energies (generally the energy changes along the reaction pathway) should be calculated with a rather high accuracy of 1 kcal/mol, which defines the required precision for computations in this field.

## Electronic Structure Methods: Accuracy vs. Computational Cost

There are numerous difficulties associated with computational studies of the reactivity of iron-containing compounds. First, the reaction of ZVI with molecules is a complex chemical process that besides the chemical changes also involves physisorption and chemisorption events. Second, the open-shell iron compounds may have several spin states. Therefore, it is necessary to identify the correct ground state and to account for the possibility of crossing between states of different multiplicities along the reaction path. It is also important to pay attention to the choice of a suitable basis set and examine the role of dynamic and non-dynamic electron correlation and relativistic effects.[2] Other complications may arise from the multireference character of certain iron containing compounds.

### *Wave Function Theory (WFT)*

The Hartree–Fock (HF) approximation (or the self-consistent field (SCF) method), which treats the wave function as a Slater determinant of one-electron functions, neglects by definition the correlation of spin-opposite electrons, i.e., the electron correlation energy. Unrestricted single reference methods give rise to spin-contamination issues when applied to open-shell systems, i.e., the unrestricted HF (UHF) wavefunction is not an eigenfunction of the total spin operator, $S^2$, and $<S^2> \geq S(S+1)$. Common consensus is that if $[<S^2>-S(S+1)]/[S(S+1)] < 10\%$, sufficiently accurate energies are obtained. On the other hand, restricted open-shell HF (ROHF) calculations with the correct $<S^2>$ can generate unphysical results due to symmetry breaking artifacts and do not allow correct spin polarization.[3] Therefore, it is sometimes necessary to use several Slater determinants or include more states for highly spin contaminated systems (especially for their transition states), i.e., to account for non-dynamic correlation using multiconfigurational SCF (MCSCF), typically complete active space (CAS) SCF.

The most widely used methods for determining the dynamic electron correlation energy are Møller–Plesset (MP) perturbation theory and coupled-clusters (CC) theory, which use a HF wave function as a reference function. The CC method with single, double and perturbative triple excitations (CCSD(T)) is currently regarded as the reference method, i.e., the "gold standard" of quantum chemistry, when applied with large basis set or complete basis set (CBS). CCSD(T) is also quite efficient at reducing UHF spin contamination to acceptable levels, meaning that the results of UHF- and ROHF-based coupled cluster calculations are often very similar. Because of the inherently multireference nature of many iron species, such single reference computations may provide inaccurate description of both non-dynamic and dynamic electron correlation, and therefore it is usually necessary to check the amplitudes in a cluster expansion (the singly and doubly excitation amplitudes in CCSD must be not larger than 0.2).[2] $T_1$ and $D_1$ diagnostics are mathematically rigorous indications of the quality of an open-shell CC wave function (using CC amplitudes). Values of $T_1 > 0.05$ or $D_1 > 0.15$ usually indicate some multireference character to the wave function.[4] Alternatively, the leading determinant of a CAS wave function must be strictly dominant (by more than 90%).[2, 4] However, the CCSD(T) method is only applicable to small systems owing to its scaling ($\sim N^7$). It should be noted that the less expensive MP perturbation theory ($\sim N^5$ for MP2) is not recommended for TM as it is sensitive to spin contamination,[3] not fully balanced for intermolecular forces (e.g. MP2 overestimates dispersion interactions) and may fail in its description of transition states.[5]

For multireference iron complexes, the multireference configuration interaction (MRCI) method with single and double excitations of MCSCF wave function is considered a benchmark method. However, its scaling is also enormous (Table 1). In addition, the Davidson correction MRCI+Q has to be used in order to obtain size consistent results. A much cheaper and less accurate alternative is to use perturbation theory over the CASSCF wave function (e.g., CASPT2). A general limitation of WFT methods is their rare implementation (see Table 1) under periodic boundary conditions (PBC).

### *Density Functional Theory (DFT)*

DFT methods are usually capable of predicting the properties of open-shell systems quite satisfactorily because they model correlation effects in a different way. A spin-polarized version of DFT can also be effectively formulated under PBC together with plane wave (PW) or localized basis sets, and the scaling of computational cost is very good ($\sim N^{3-4}$). However, widely used classical DFT functionals, i.e., the local density approximation (LDA) and generalized gradient approximation (GGA), are impaired by inherent limitations. The self-interaction error (SIE) and neglect of non-local electron correlation effects are two of the most serious limitations in using these functionals to model molecule–nZVI interactions and processes. Thus, adsorption energies and reaction barriers for such systems cannot yet be calculated with chemical accuracy. The SIE can be effectively reduced by using hybrid DFT functionals involving some portion of the exact HF exchange. Moreover, short-range functionals, such as the screened hybrid functional HSE06,[6] are an effective alternative for periodic systems. Non-local correlations represent a more difficult problem. Nevertheless, in recent years, empirically corrected (DFT-D) and approximate non-local density functionals (vdW-DF) have been developed.[7]

The random phase approximation (RPA)[8] to the correlation energy (as a "post-DFT" method) represents another advance. It naturally includes non-local electron correlation with the correct asymptotic decay. Besides improved description of bulk material properties, RPA also corrects physisorption and chemisorption energies. The computational costs of RPA in the case of periodic surface calculations are quite high, with $\sim N^6$ scaling and a large memory required.

| Table 1. Comparison of different electronic structure methods | | | | | |
|---|---|---|---|---|---|
| Method | Scaling[a] | Size[b] | DynC[c] | NonDC[d] | PBC[e] |
| HF | $N^{3-4}$ | 50-100 | | | √ |
| MP2 | $N^5$ | 20-30 | * | | √ |
| MP4 | $N^7$ | 10-20 | ** | | |
| DFT | $N^{3-4}$ | 50-200 | * | | √ |
| RPA | $N^6$ | 10-30 | ** | | |
| CCSD | $N^6$ | 10-30 | ** | | |
| CCSD(T) | $N^7$ | 10-30 | *** | | |
| CCSDT | $N^8$ | 5-15 | *** | | |
| MCSCF | $n \times N^4$ | 15-25 | * | *** | |
| MRCI | $n \times N^6$ | <10 | *** | *** | |
| QMC | $N^{3-4}$ | <250 | *** | *** | √ |
| Full CI | $N!$ | <5 | Exact | Exact | |
| [a] Order of the computational scaling with the number of basis functions ($N$) and the dimension of the multireference space ($n$). [b] Size range for which the methods are typically applied in the literature. [c] Degree of treatment of dynamic electron correlation: 0 asterisks (not included) – 3 asterisks (satisfactory treatment). [d] Non-dynamic electron correlation. [e] Implementation under PBC. Data partially reprinted from Ref. [9] with permission from Elsevier. | | | | | |

### *Other Methods*

Finally, in addition to the methods that are solved with finite basis sets, stochastic approaches can be used to solve the Schrödinger equation, namely variational and diffusion quantum Monte Carlo (QMC) methods. Both these methods aim to provide an exact solution (competitive to full CI) of the Schrödinger equation (with only limitations from the fixed node approximation or pseudopotential, if used).[10] QMC scaling is favorable (~$N^{3-4}$), there are very small demands on memory in diffusion QMC and almost ideal scaling with the number of processors. Therefore, QMC overcomes many of the limitations associated with current WFT and DFT methods and provides reliable energy differences for rather complex molecular systems[11] or surfaces owing to its implementation under PBC. However, due to the large "pre-factor" within the $N^{3-4}$, these methods are currently considered too computationally expensive. Nevertheless, they provide benchmarks not easily accessible by any other method.

## Models of ZVI Nanoparticles

To calculate chemical changes of a nanoparticle three different models can in principle be used: atomic, cluster and surface models. One limiting model, the atomic model (Figure 1a), is useful for (i) obtaining accurate $\Delta G^{\ddagger}$, and (ii) identifying elementary reaction mechanisms, and as such represents the simplest model for nZVI reactions, allowing application of coupled cluster and high level multireference calculations. However, such a simple model cannot fully describe the complexity of a metal nanoparticle.

Iron clusters may be considered a logical and quite realistic model of nZVI. However, they are still rather small compared to a real nanoparticle. In particular, symmetric (icosahedral) clusters of "magic" sizes as $Fe_{13}$ (~0.5 nm particle size), $Fe_{55}$ (~1 nm), $Fe_{147}$ (~1.4 nm), $Fe_{309}$ (~1.8 nm) or $Fe_{561}$ (~2.3 nm) are popular choices for modeling nanoparticles[12, 13] because of their reduced number of possible high-symmetric absorption sites (Figure 1a). However, the methods available are limited due to number of atoms and GGA DFT is typically applied, providing rather qualitative predictions for iron.

The other limiting model of a nanoparticle is as a solid surface. In this particular case, one can benefit from PBC, which allows rather advanced and reliable methods like RPA and QMC to be used.[10, 14] On the other hand, one has to consider the reaction on various surfaces.

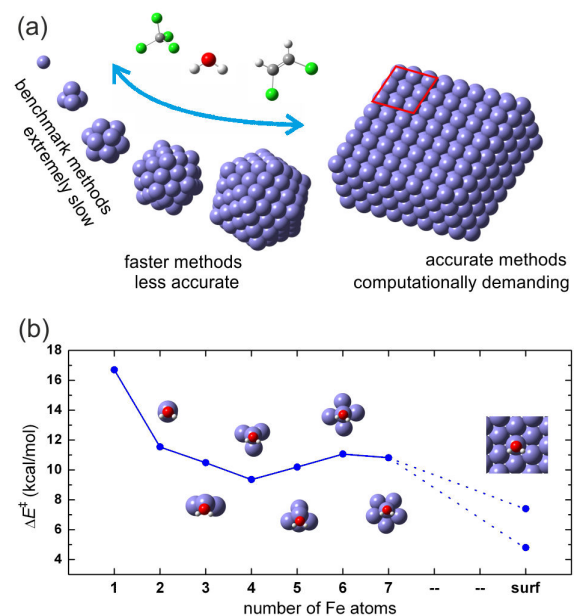

Figure 1. (a) Models of iron nanoparticle. (b) Change of activation energy $\Delta E^{\ddagger}$ for $Fe_n + H_2O$ reaction with respect to size of cluster at PW91 level. Adsorption geometries are shown.

## Current State-of-the-Art for nZVI Reactions

### Reaction of Iron with Water

The reaction of iron with water under anaerobic conditions, also known as anaerobic corrosion, is of wide importance. In our recent studies,[5, 14-16] we used several methods and models to study the reaction. First, we used the atomic model, for which advanced and highly accurate WFT methods are applicable and may provide reference data, to compare with the performance of other methods. The reaction of atomic iron with water Fe + 2($H_2O$) → Fe(OH)$_2$ + $H_2$ was used as a model system in order to decipher the elementary reaction mechanism. The free energy profiles showed two separated single electron reaction steps (Figure 2a): Fe$^0$ + $H_2O$ → HFe$^I$OH, followed by HFe$^I$OH + $H_2O$ → Fe$^{II}$(OH)$_2$ + $H_2$. For all intermediates on the reaction path, the quintet state was favorable and the CCSD(T)-3s3p-DKH/CBS level was adequate due to the single reference character of all considered species.[14] The first reaction step, i.e., splitting of the H-OH bond and formation of the HFeOH molecule,[17] was found to be the rate limiting step, with a reaction barrier ($\Delta G^{\ddagger}_{298K}$ = 29.2 kcal/mol)[5, 14] almost two fold higher than the activation barrier of the second step, in which Fe(OH)$_2$ and hydrogen are generated. To assess the accuracy of various approaches, we compared reaction profiles obtained with the GGA (PBE), short-range hybrid (HSE06), hybrid (B97-1) and the RPA methods against the CCSD(T)-3s3p-DKH/CBS profile (Figure2a) and observed the following systematic increase in the accuracy of the density functionals: GGA < hybrid < RPA (activation energy $\Delta E^{\ddagger}$ for the rate limiting step of 16.7, 24.7 and 31.1 kcal/mol, respectively), where the latter approach provided good agreement with the CCSD(T)-3s3p-DKH/CBS reaction profile (Table2, Figure2a).

Reaction of an iron surface (100, 110, 111) with water was also evaluated.[14, 15, 18, 19] It was shown that the reaction mechanism was in many respects analogous to the reaction of water with a single Fe atom. The rate limiting step was again breakage of the H-OH bond (Figure 2a). The results revealed that local functionals vastly underestimate activation barriers, even for less localized states of a metallic solid: the barrier height calculated for water dissociation on the Fe(100) surface by RPA was found to be 14.6 kcal/mol (similar to HSE06 value, Table 2), whereas a value of 7.4 kcal/mol was obtained with the PW91 functional.[14] Interestingly, the barrier from RPA was significantly lower than that obtained for the reaction with an Fe atom. Therefore, for the purpose of this work, we performed additional DFT GGA calculations for the reaction Fe$_n$ + $H_2O$ (n = 2-7) (Figure 1a) and showed that the activation energy decreased as the number of iron atoms was increased, with the largest change observed on going from atom to dimer.

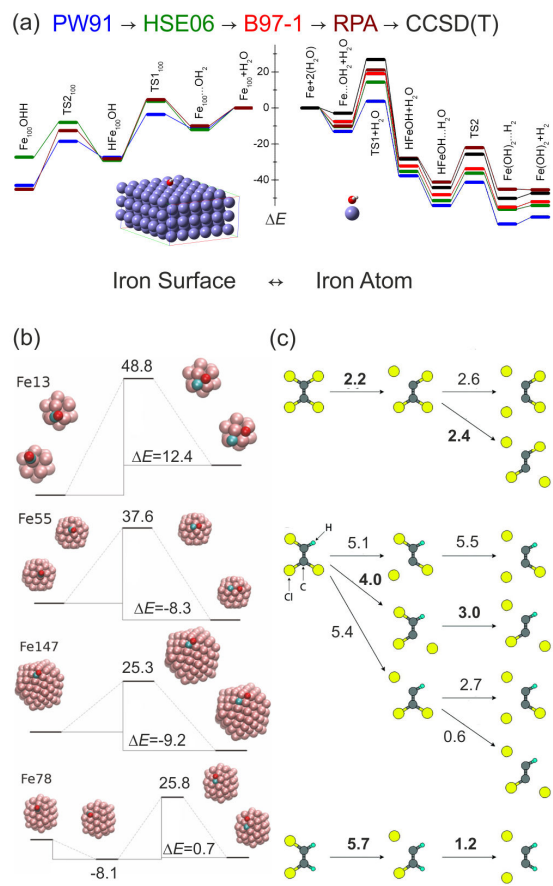

Figure 2. Minimal energy pathways (kcal/mol) for dissociation of various molecules on iron. (a) $H_2O$ on an Fe atom and Fe(100) surface using various methods. Reprinted with permission from Ref. 14. Copyright 2013, American Chemical Society. (b) CO on Fe$_n$, n=13, 55, 147, and 78. Reprinted with permission from Ref. 12. Copyright 2013, AIP Publishing LCC. (c) PCE, TCE and cis-DCE on an Fe(110) surface. Reprinted with permission from Ref. [20]. Copyright 2009, American Chemical Society.

### Reaction of Iron with NH$_3$, CO and CO$_2$

The bonding and dissociation of a NH$_3$ molecule and its fragments were studied on an icosahedral Fe$_{55}$ cluster,[21] Fe(111),[22] and Fe(110)[23] surfaces. The reaction mechanism NH$_3$ → NH$_2$ + H → NH + H + H → N + H + H + H was the same for all models and similar activation energies were also obtained. The rate limiting step was dissociation of NH$_3$ for reaction on the cluster, dissociation of NH$_2$ for Fe(111), and NH for the Fe(110) surface, with activation barriers of $\Delta E^\ddagger$ = 34.1, 28.5, or 26.7 kcal/mol, respectively.

The reactivity of 0.5-1.4 nm iron nanoparticles with CO and corresponding bulk surfaces has been systematically studied using GGA on ideally symmetric (Fe$_{13}$, Fe$_{55}$, and Fe$_{147}$) and more realistic rugged (Fe$_{78}$) nanoparticles.[12] The activation energies for CO dissociation were found to vary between 25 and 49 kcal/mol (Figure 2b). Increasing the particle size and roughness resulted in lower activation energies. For a single particle, variations as large as 20.7 kcal/mol were observed due to different adsorption sites and pathways, demonstrating the importance of local particle morphology. Studies on the smallest particles and bulk surfaces (Fe(111) using various lattice constants) showed that these systems were possibly not reliable as models for catalysis on larger iron nanoparticles (Table 2). The differences in reaction mechanisms can be rationalized by the varying Fe-Fe bond lengths in different particles, leading to changes in back-bonding between the iron surface and CO.

For the sake of completeness, we note that reactions of iron with CO$_2$ were also studied, but the activation energies obtained at GGA level were not very consistent for Fe(100) surfaces (5 kcal/mol[24] vs. 27 kcal/mol[25]). The activation barrier predicted from the atomic model was 23.4 kcal/mol.[26] The observed inconsistencies call for further research in this sphere and application more elaborate theoretical methods.

Table 2. Activation energies $\Delta E^\ddagger$ (kcal/mol) for reactions of various molecules with iron using atomic, cluster and surface models for ZVI.

| Reaction | GGA DFT | Hybrid DFT | Best Method | Ref. |
|---|---|---|---|---|
| Fe+H$_2$O | 16.7 | 24.7,[b] 26.5[c] | 29.7,[e] 31.1[f] | 5, 14 |
| Fe+CO$_2$ | --- | 23.4[d] | --- | 26 |
| HFeOH+CCl$_4$ | --- | 6.4[d] | 23.8[g] | 27 |
| FeCl$_2$+CCl$_4$ | --- | --- | 21.1[g] | 27 |
| Fe$_{2-7}$+H$_2$O | 9.4-11.5 | --- | --- | This w. |
| Fe$_2$+H$_2$O | --- | 6.3[d] | --- | 28 |
| Fe$_{55}$+NH$_3$ | 34.1[a] | --- | --- | 21 |
| Fe$_{13}$+CO | 48.8 | --- | --- | 12 |
| Fe$_{55}$+CO | 37.6 | --- | --- | 12 |
| Fe$_{78}$+CO | 25.8 | --- | --- | 12 |
| Fe$_{147}$+CO | 25.3 | --- | --- | 12 |
| Fe(100)+H$_2$O | 7.4, 8.1 | 15.7[b] | 14.6[f] | 14, 15, 19 |
| Fe(111)+H$_2$O | 4.8 | 13.3[b] | --- | 15 |
| Fe(110)+H$_2$O | 4.4 | --- | --- | 18 |
| Fe(111)+NH$_3$ | 28.5[a] | --- | --- | 22 |
| Fe(110)+NH$_3$ | 26.7 | --- | --- | 23 |
| Fe(100)+CO | 30.9-47.7 | --- | --- | 12 |
| Fe(100)+CO$_2$ | 5.0, 27.1 | --- | --- | 24, 25 |
| Fe(110)+PCE | 2.4 | --- | --- | 20 |
| Fe(110)+TCE | 4.0 | --- | --- | 20 |
| Fe(110)+DCE | 5.7 | --- | --- | 20 |

[a] rPBE. [b] HSE06. [c] B97-1. [d] B3LYP. [e] CCSD(T)-sp-DKH/CBS, $\Delta G^\ddagger_{298K}$=29.2 kcal/mol. [f] RPA. [g] CCSD(T)/TZV.

### Pollutants: Chlorinated Ethenes and Methanes

Periodic GGA has been used to study the gas-phase dissociation of perchloroethene (PCE), trichloroethene (TCE) and cis-dichloroethene (cis-DCE) on Fe(110).[20] The activation energies were found to decrease as the chlorination number increased (Figure 2c). The rate limiting step for PCE dissociation was the second chlorine cleavage, whereas the rate limiting step for TCE and cis-DCE was the first chlorine cleavage. The activation energies of the rate limiting steps for PCE, TCE and cis-DCE were 2.4, 4.0, and 5.7 kcal/mol, respectively.[20] The relative gas-phase reactivity order was found to be PCE > TCE > cis-DCE: at room temperature (300 K), the PCE dechlorination rate was 14 and 338 times faster than that of TCE and cis-DCE, respectively.

Because reaction of iron atoms with water may produce HFeOH,[14, 17, 29] reaction of CCl$_4$ with HFeOH was studied at the atomic level using the CC method and Marcus-Hush theory. An activation energy of 23.8 kcal/mol for reaction HFeOH + CCl$_4$ → HFeClOH + ·CCl$_3$ was obtained at CCSD(T) level.[27] Calculations also show that the corresponding transition state arises from crossing of electronic states in which the configuration of Fe changes from a quintet high spin state in the $Fe^{II}$ reactant to a sextet high spin state in the $Fe^{III}$ products. The magnitude of the reaction barrier was consistent with absence of products in the atom-dropping experiments under low temperatures.[29]

## Summary and Outlook

The modeling of nZVI particles as iron clusters is particularly promising because this approach can include all the potential complexity of nanoparticles. However, the large number of iron atoms included (still small with respect to an actual iron nanoparticle) dictates that rather low accuracy methods have to be used. Hence, the predicted reaction barriers (typically from GGA DFT) often are not comparable with experimental values. Therefore, besides of hybrid DFT modeling, it may be preferable to model molecular reactions with iron atoms to obtain accurate reaction barriers using high-level methods. In addition, we have shown on several examples that, surprisingly, the single reference CCSD(T) method can be used if one carefully checks cluster expansion. Using an atomic model, it is also easy to identify which physical phenomena (e.g. relativistic effects, static and dynamic electron correlation effects, non-local correlation effects) has to be explicitly taken into account and which can be safely neglected.

However, extension of the model to a few iron atoms is very difficult. Even an iron dimer has a strong multireference character[30] and the best multireference methods, such as MRCI, are only properly applicable for up to three or four iron atoms. Hopefully, rapid development of QMC methods in the future may provide more accurate energies at a computationally affordable cost. In addition, it is possible to overcome the principal limitation of extending atomic models to several iron atoms by using a different direction for the periodic cell, i.e., using surface models. Reactions on iron surfaces enable some collective properties to be taken into account. In addition, smaller number of atoms in the cell may be favorably utilized by the option for using highly accurate methods, such as QMC or RPA under PBC. Also the recent implementation of CC method for PBC has demonstrated considerable promise.[31] Hybrid DFT functionals represents a computationally less demanding alternative which seems quite reliable in surface chemistry,[14] however more benchmark calculations are needed. New experimental data on the reactivity of nZVI may also help to identify shortcomings of current methods and models.

Understanding reactivity of nZVI and iron, which is the most abundant element on the Earth, requires investigation of rather complex surface or cluster models. It also calls for application of elaborate methods of quantum chemistry. In a near future, one may benefit from a brutal force approach[13] or application of more sophisticated approaches, e.g., hybrid QM/QM methods. Overall, despite recent progress, the theoretical description of nZVI and its reactivity is still challenging.

## Acknowledgments

Financial support through projects P208/11/P463, CZ.1.05/2.1.00/03.0058, CZ.1.07/2.3.00/20.0017, and TE01010218 is gratefully acknowledged.

**Keywords:** nano zero-valent iron, reactivity, DFT, *ab-initio* calculations, reaction barrier